\documentclass[submission, Phys]{SciPost}

\pdfoutput=1
\usepackage{mathtools}
\usepackage[utf8]{inputenc} 
\usepackage[T1]{fontenc} 	
\usepackage[english]{babel} 

\usepackage[bitstream-charter]{mathdesign}
\usepackage{geometry} 		
\usepackage{amsmath} 		
\usepackage{mathtools} 		
\usepackage{float} 			
\usepackage{graphicx} 		
\usepackage{tabularx} 		
\usepackage{booktabs} 		
\usepackage{color, xcolor} 	
\usepackage{pdfpages} 		
\usepackage{extarrows} 		
\usepackage{multirow} 		
\usepackage{multicol} 		
\usepackage{caption} 		
\usepackage{comment}
\usepackage{subcaption} 	
\usepackage{enumitem} 		
\usepackage{xspace} 		
\usepackage{stackrel} 		
\usepackage{tikz} 			
\usetikzlibrary{calc}
\usepackage{braket} 		
\usepackage{bm} 			
\usepackage{tensor} 		
\usepackage{slashed} 		
\usepackage{siunitx} 
\usepackage{lastpage} 		
\usepackage{cite} 			
\usepackage[normalem]{ulem} 
\usepackage{fontawesome} 	
\usepackage{tocloft} 		
\usepackage{titlesec} 		
\usepackage{doi} 			
\usepackage[most]{tcolorbox} 					
\usepackage[capitalize]{cleveref} 	
\usepackage[nottoc, notlot, notlof]{tocbibind} 	
\usepackage[ruled, vlined]{algorithm2e} 		
\usepackage{makecell}
\usepackage{pifont}
\usepackage[dvipsnames]{xcolor}

\hypersetup{
	pdftitle={NAE, Statistically},
	pdfauthor={Ranit Das, Jonathan Ostertag-Henning, Tilman Plehn, and Lorenz Vogel},
    linktoc=section,
	breaklinks=true,			
	colorlinks=true, 			
	linkcolor={red!50!black}, 	
	citecolor={blue!50!black}, 	
	urlcolor={blue!80!black} 	
} 

\DeclareSymbolFont{usualmathcal}{OMS}{cmsy}{m}{n}
\DeclareSymbolFontAlphabet{\mathcal}{usualmathcal}

\SetArgSty{textnormal}
\SetKwComment{Comment}{{\small\#}~}{}
\SetCommentSty{mycommfont}

\setitemize{itemsep=2pt, parsep=0pt} 				
\setenumerate{itemsep=2pt, parsep=0pt} 				
\crefname{section}{Section}{Sections}
\crefname{equation}{Eq.}{Eqs.}
\crefname{figure}{Figure}{Figures}
\crefname{table}{Table}{Tables}
\Crefname{section}{Section}{Sections}
\Crefname{equation}{Equation}{Equations}
\Crefname{figure}{Figure}{Figures}
\Crefname{table}{Table}{Tables}
\newcommand{\ie}{i.e.\@\xspace} 	

\newcommand{\eqc}{\;\text{,}} 		
\newcommand{\eqp}{\;\text{.}} 		

\newcommand{\cmark}{\ding{51}}
\newcommand{\xmark}{\ding{55}}

\newcommand{\class}{C}

\newcommand{\Langle}{\bigl\langle}
\newcommand{\Rangle}{\bigr\rangle}
\newcommand{\XLangle}{\Bigl\langle}
\newcommand{\XRangle}{\Bigr\rangle}
\newcommand{\XXLangle}{\biggl\langle}
\newcommand{\XXRangle}{\biggr\rangle}

\newcommand{\mwith}{\text{with}}
\newcommand{\mand}{\text{and}}

\newcommand{\qqquad}{\qquad\quad}
\newcommand{\qqqquad}{\qquad\qquad}

\newcommand{\dd}{\mathop{}\!\text{d}}
\newcommand{\real}{\mathbb{R}} 				
\newcommand{\var}{\operatorname{Var}} 		

\newcommand{\order}{\mathcal{O}} 			
\newcommand{\range}[2]{#1\:\!\ldots\:\!#2}	
\newcommand{\really}{\stackrel{!}{=}}
\newcommand{\mean}[1]{\left\langle#1\right\rangle}

\newcommand{\pt}{p_{\text{T}}} 	
\newcommand{\pti}{p_{\text{T},i}} 
\newcommand{\ptjet}{p_{\text{T},\,\text{jet}}}
\newcommand{\etajet}{\eta_{\,\text{jet}}}

\newcommand{\kt}{k_{\text{T}}}
\newcommand{\loss}{\mathcal{L}} 	

\newcommand{\gauss}{\mathcal{N}} 	

\newcommand{\kl}[2]{D_{\text{KL}}[#1,#2]}

\newcommand{\etaphi}{$\eta$--$\phi$\xspace} 

\newcommand{\pytorch}{\texttt{PyTorch}\xspace}

\newcommand{\adam}{\texttt{Adam}\xspace}

\newcommand{\pythia}{\textsc{Pythia}\xspace}

\newcommand{\delphes}{\textsc{Delphes}\xspace}
\newcommand{\fastjet}{\textsc{FastJet}\xspace}

\newcommand{\toptagging}{\textsc{TopTagging}\xspace}

\newcommand{\arXiv}[2][]{%
	\ifthenelse{\equal{#1}{}}%
	{\href{http://arxiv.org/abs/#2}{arXiv:#2}}%
	{\href{http://arxiv.org/abs/#2}{arXiv:#2~[#1]}}}

\newcommand{\gev}{\text{GeV}}
\newcommand{\tev}{\text{TeV}}

\def\slashchar#1{\setbox0=\hbox{$#1$}           
   \dimen0=\wd0                                 
   \setbox1=\hbox{/} \dimen1=\wd1               
   \ifdim\dimen0>\dimen1                        
      \rlap{\hbox to \dimen0{\hfil/\hfil}}      
      #1                                        
   \else                                        
      \rlap{\hbox to \dimen1{\hfil$#1$\hfil}}   
      /                                         
   \fi}

\newcommand{\tikznode}[2]{%
\ifmmode%
\tikz[remember picture,baseline=(#1.base),inner sep=0pt] \node (#1) {$#2$};%
\else
\tikz[remember picture,baseline=(#1.base),inner sep=0pt] \node (#1) {#2};%
\fi}

\def\mathswitchr#1{\relax\ifmmode{\text{#1}}\else$\text{#1}$\xspace\fi}
\def\mathswitch#1{\relax\ifmmode#1\else$#1$\xspace\fi}

\newcommand{\pdata}{p_{\text{data}}}
\newcommand{\ptrue}{p_{\text{true}}}
\newcommand{\sigstat}{\sigma_{\text{stat}}}
\newcommand{\ptheta}{p_{\theta}}
\newcommand{\etheta}{E_{\theta}}
\newcommand{\ttheta}{T_{\theta}}
\newcommand{\ctheta}{C_{\theta}}
\newcommand{\ztheta}{Z_{\theta}}

\newcommand{\rocauc}{\text{AUC}}
\newcommand{\sigeff}{\epsilon_{\text{sig}}}
\newcommand{\bkgrej}{\epsilon_{\text{bkg}}^{-1}}

\newcommand{\dtrain}{\mathcal{D}_{\text{train}}}
\newcommand{\dtest}{\mathcal{D}_{\text{test}}}
\newcommand{\dgrid}{\mathcal{D}_{\text{grid}}}

\graphicspath{{./figures/}}

\begin{document}

\begin{center}
    {\Large\textbf{\color{scipostdeepblue}{NAE, Statistically}}}
\end{center}

\begin{center}\textbf{
    Ranit~Das\textsuperscript{1},
    Jonathan~Ostertag-Henning\textsuperscript{1},
    Tilman~Plehn\textsuperscript{1,2}, and
    Lorenz~Vogel\textsuperscript{1}
}\end{center}

\begin{center}
    {\bf 1} Institut f\"{u}r Theoretische Physik, Universit\"{a}t Heidelberg, Germany\\
    {\bf 2} Interdisciplinary Center for Scientific Computing (IWR), Universit\"{a}t Heidelberg, Germany
\end{center}

\begin{center}
    \today
\end{center}
 
\section*{\color{scipostdeepblue}{Abstract}}
\textbf{\boldmath{Searches for new physics using neural anomaly scores have transformative potential, but suffer from a lack of statistical interpretability. The normalized autoencoder (NAE) provides a probabilistic interpretation of the standard bottleneck architecture, tying the anomaly score to a learned likelihood. We validate this relation for a toy model, test it for jets using a dual-NAE setup, and show how a Bayesian NAE learns this likelihood with an uncertainty.}}

\vspace{10pt}
\noindent\rule{\textwidth}{1pt}
\tableofcontents\thispagestyle{fancy}
\noindent\rule{\textwidth}{1pt}
\vspace{10pt}

\clearpage
\section{Introduction}
\label{sec:intro}

The defining goal of the LHC program is the search for physics beyond the Standard Model (BSM). Traditionally, BSM searches at the LHC have been theory-driven, starting with a proposed model and its experimental signatures and then comparing these signatures to data. This method led to the discovery of the top quark and the Higgs boson, but requires theory predictions that guess the correct model in detail. A more modern approach starts with a precision analysis of a physics-motivated kinematic regime and then interprets these BSM-independent precision measurements with a range of established BSM model simulations.

If we have doubts which correct BSM model to test with simulation-based inference, we can use anomaly searches~\cite{Kasieczka:2021xcg,Karagiorgi:2021ngt,Belis:2023mqs,Amram:2026vkc} at least as a guide. Such anomaly searches are enabled by modern machine learning~\cite{Kramer:2026eoq,Plehn:2022ftl}. Weakly supervised anomaly searches make a model assumption by defining a narrow signal region in phase space, but they extract the SM expectation from data. In current realizations~\cite{ATLAS:2023azi,CMS:2025sch}, signal regions are defined based on a simple observable, like an invariant mass of two particles~\cite{Nachman:2020lpy,Hallin:2021wme,Hallin:2022eoq,Das:2023bcj,Sengupta:2023xqy,Sengupta:2023vtm,Das:2024eie,Oleksiyuk:2025pmu,Brennan:2025fqy,Breso-Pla:2026tlz, Cheng:2026iiu, Mikuni:2026ced, Das:2024fwo, Das:2026spe}. This way, they generalize bump hunts by accounting for correlations with the remaining phase space directions. Statistically, such analyses are well-defined, provided we can account for the look-elsewhere effect of an automatic scan through many signal regions. 

The ultimate goal for model-agnostic BSM searches is to flag events as anomalous in relation to the majority of observed events in terms of a well-defined score. Unsupervised anomaly detection looks for rare events that are different from the bulk of the data, without committing to a BSM signal hypothesis. We emphasize that given the LHC data stream the notion of out-of-distribution events is not justified. ML-based unsupervised anomaly detection approaches rely on bottleneck-based autoencoders~\cite{Heimel:2018mkt,Farina:2018fyg,Finke:2021sdf,Dillon:2021nxw,Dillon:2022mkq,Ngairangbam:2025fst,CMS:2025lmn,Eble:2024tpr} or on normalizing flows~\cite{Buss:2022lxw,Krause:2023uww,Patel:2026npj}. Anomaly scores can be enhanced using physics information~\cite{Ostdiek:2021bem,Hao:2022zns,Dillon:2023zac,Metodiev:2023izu,Favaro:2023xdl,Gandrakota:2024ruu,Li:2026azw, Jia:2026pes}, such that they can be used to search for broad classes of BSM models, where simulation-based inference is hampered by high-dimensional model parameter spaces~\cite{Barron:2021btf}.

Although unsupervised anomaly detection is technically feasible, its application in LHC analyses is limited by the absence of a statistical interpretation~\cite{Caron:2021wmq,Mikuni:2023tok,Craig:2024rlv,Hein:2025uhj,Araz:2026img}. In this paper, we ask two questions:
\begin{enumerate}
    \item What is the statistical interpretation of the anomaly score of a normalized autoencoder? 
    \item Can we extract it with an uncertainty estimate?
\end{enumerate}
We start with a discussion of the normalized autoencoder, the relation of its training objective with the underlying likelihood, and a Bayesian network setup of the NAE in \cref{sec:bnae}. In \cref{sec:toy} we introduce and analyze a toy model for which we know the correct likelihood. We find that the (B)NAE indeed learns the correct likelihood, ideally with a learned temperature. For realistic LHC applications, we do not know the correct likelihood, so in \cref{sec:toy_lr_test} we introduce a dual-NAE test, where we compare the outputs of two complementary NAEs to the likelihood ratio from a supervised classifier. 

In \cref{sec:jets}, we train the (B)NAE on the top-tagging dataset also used in Ref.~\cite{Dillon:2022mkq}. We study the anomaly detection performance and its Bayesian uncertainties in \cref{sec:jets_performance}. In \cref{sec:jets_lr_test}, we apply the dual-(B)NAE test to jet images, using a calibrated supervised classifier as a reference for the unknown likelihood ratio. We find that the learned energy differences approximately reproduce the classifier log-likelihood ratio. Appendix~\ref{app:jets_details} details our data preprocessing. Appendix~\ref{app:hyperparameters} provides additional training details and hyperparameters. Appendix~\ref{app:results} contains supplementary results on uncertainty calibration, (B)NAE-generated samples, and BNAE-learned energies.

\clearpage
\section{Bayesian normalized autoencoder}
\label{sec:bnae}

An energy-based network assigns a scalar energy to each phase-space point,
\begin{align}
    \etheta(x): \real^{d_{x}} \to \real \eqp
\end{align}
The energy is a learned mapping and related to a learned probability density through the Boltzmann distribution
\begin{align}
    \ptheta(x) = \frac{e^{-\etheta(x)/T}}{\ztheta}
    \qquad\mwith\quad
    \ztheta = \int \dd x\: e^{-\etheta(x)/T} \eqp
    \label{eq:boltzmann}
\end{align}
In many applications, the temperature is ignored by setting $T = 1$, but we can fix it to a different value or even make it learnable. The partition function $\ztheta$ ensures that $\ptheta(x)$ is properly normalized over $x$. The network is trained by minimizing the negative log-likelihood,
\begin{align}
    \loss(x\vert\theta) 
    &= -\log\ptheta(x) = \frac{\etheta(x)}{T} + \log\ztheta \notag \\
    \Rightarrow\qqquad
    \loss(\theta) 
    &= \XXLangle \frac{\etheta(x)}{T} + \log \ztheta \XXRangle_{\pdata} \eqp
    \label{eq:ebm_loss}
\end{align}
Minimizing this loss reduces the energy on the data while accounting for the normalization of the full model distribution. Its gradient with respect to the network parameters is
\begin{align}
    \Langle \nabla_{\theta} \loss(x\vert\theta) \Rangle_{\pdata}
    = \frac{1}{T} \Langle \nabla_{\theta} \etheta(x) \Rangle_{\pdata} - 
    \frac{1}{T} \Langle \nabla_{\theta} \etheta(x) \Rangle_{\ptheta} \eqp
    \label{eq:training_ebm}
\end{align}
The first term decreases the energy on the training data, while the second term increases the energy on samples drawn from the energy-based network. They are referred to as positive and negative energies respectively. The samples drawn from the energy-based network are referred to as negative-energy samples.

To see that an energy-based network recovers the target density, we look at 
\begin{align}
    \loss[\etheta]
    &= \Langle -\log\ptheta(x) \Rangle_{\pdata} \notag \\
    &= \frac{1}{T} \int \dd x\: \pdata(x) \etheta(x) + \log\ztheta \eqp
\end{align}
The variation of the first term is straightforward, but in the second term $\ztheta$ depends on $\etheta(x)$,
\begin{align}
    0 \really \delta \loss[\etheta]
    &= \frac{1}{T} \int \dd x\: \pdata(x) \delta\etheta(x)
    + \frac{\delta\ztheta}{\ztheta} \notag \\
    &= \frac{1}{T} \int \dd x\: \pdata(x) \delta\etheta(x) + \frac{1}{\ztheta}
    \int \dd x\: \frac{\dd e^{-\etheta(x)/T}}{\dd \etheta(x)} \delta\etheta(x) \notag \\
    &= \frac{1}{T} \int \dd x\: [ \pdata(x) - \ptheta(x) ] \delta\etheta(x) \eqp
\end{align}
A minimum in the loss implies
\begin{align}
    \ptheta(x) = \pdata(x) \eqc
    \label{eq:learn-ebm}
\end{align}
so the trained energy-based network reproduces the training data density by adjusting the scalar mapping $\etheta$. This relation makes it clear that a standard autoencoder without a partition function does not provide a link to the likelihood. The problem with this conceptual strength of energy-based networks is that the second term in Eq.\eqref{eq:training_ebm} is hard to evaluate in high dimensions.

\subsubsection*{Normalized autoencoder}

An autoencoder (AE) reconstructs an input $x \in \real^{d_{x}}$ via a low-dimensional latent representation $z \in \real^{d_{z}}$,
\begin{align}
    x \; \xmapsto{\;f_\text{e}(x)\;} \; 
    z \; \xmapsto{\;f_\text{d}(z)\;} \; x^{\prime} \eqp
\end{align}
The encoder $f_{\text{e}}(x)$ and decoder $f_{\text{d}}(z)$ are trained by minimizing the MSE reconstruction loss,
\begin{align}
    \loss_\text{MSE}
    = \Langle \vert x-x^{\prime} \vert^{2} \Rangle_{\pdata} \eqp
\end{align}
Without the constraint of a low-dimensional latent representation, this is trivially solved by the identity. The information bottleneck $d_{z}\ll d_{x}$ forces the AE to learn a compressed representation of the training data $\pdata$~\cite{Blance:2019ibf,Vent:2025ddm}. The value of the reconstruction loss can be used as an anomaly score, but without a proper statistical interpretation. 

The problem with the MSE anomaly score is that for signals and backgrounds with different intrinsic dimensionality it is dominated by the trivial compressibilities of these hypotheses. The MSE will separate an intrinsically high-dimensional (more complex) signal from a low-dimensional (less complex) background, but not vice versa~\cite{yoonAutoencodingNormalizationConstraints2023}. At the LHC the MSE identifies top jets hidden among QCD jets or QCD jets hidden among semi-visible jets, but not the other way around~\cite{Finke:2021sdf}. 

The normalized autoencoder (NAE)~\cite{yoonAutoencodingNormalizationConstraints2023, songHowTrainYour2021, Dillon:2022mkq, CMS:2025lmn} identifies the reconstruction error with the energy,
\begin{align}
    \etheta(x) = \vert x - x^{\prime} \vert^{2} \eqc
    \label{eq:energy}
\end{align}
where $\theta$ are the network weights. It is trained by minimizing the negative log-likelihood in Eq.\eqref{eq:ebm_loss}. A practical way to obtain samples from the model distribution $\ptheta(x)$ to compute the gradient in Eq.\eqref{eq:training_ebm} is through Langevin Monte Carlo (LMC) with the update rule in feature space
\begin{align}
    x_{t+1} = x_{t} + \lambda \nabla_{x} \log\ptheta(x_{t})
    + \sigma \epsilon_{t} 
    \qquad\mwith\quad 
    \epsilon_{t} \sim \gauss(0, \mathbb{1}) \eqp
    \label{eq:lmc}
\end{align}
Here, $\lambda$ is the step size and $\sigma$ is the noise standard deviation. Since
\begin{align}
    \nabla_{x} \log\ptheta(x) = -\frac{1}{T} \nabla_{x} \etheta(x) \eqc
\end{align}
the chain drifts toward low-energy regions, while the noise term allows exploration. We can choose $\lambda$ and $\sigma$ to place more weight on the gradient term than on the noise term~\cite{Dillon:2022mkq}. To converge more quickly, we then sample from the distribution at a low effective temperature, 
\begin{align}
    T_{\text{eff}} = \frac{\sigma^{2}}{2\lambda} T \eqp
    \label{eq:eff_temperature}
\end{align}
To produce reliable samples from these chains, the NAE uses on-manifold initialization. As the autoencoder compresses the data into a low-dimensional $z \in \real^{d_{z}}$, the learned decoder output follows the data manifold as a high-density region~\cite{yoonAutoencodingNormalizationConstraints2023}. On-manifold initialization first samples in the latent space and then maps the result to feature space. With an energy over latent space
\begin{align}
    \tilde{E}_{\theta}(z) = \etheta(f_{\text{d}}(z)) 
    \qquad\mand\qquad 
    \tilde{p}_{\theta}(z) =
    \frac{e^{-\tilde{E}_{\theta}(z)/\tilde{T}}}{\tilde{Z}_{\theta}} \eqc
\end{align}
where $\tilde{p}_{\theta}(z)$ is the latent-space density and $\tilde{Z}_{\theta}$ is the latent-space partition function. The sampler builds a short latent Langevin chain,
\begin{align}
    z_{t+1} = z_{t} + \tilde{\lambda} \nabla_{z} \log \tilde{p}_{\theta}(z_{t}) + \tilde{\sigma} \epsilon_{t}
    \qquad\mwith\quad 
    \epsilon_{t} \sim \gauss(0, \mathbb{1}) \eqp
\end{align}
Its final point $z_{\tilde{\tau}}$ is decoded and used as the starting point $x_{0} = f_{\text{d}}(z_{\tilde{\tau}})$ of the feature-space Langevin chain in Eq.\eqref{eq:lmc}. This provides us with network samples for the likelihood gradient while avoiding the need to explore the full feature space. 

For the jet-tagging setup in \cref{sec:jets}, the latent-space parametrization maps the variables to a compact hypersphere $\mathbb{S}^{d_{z}-1}$, allowing uniform initialization in latent space. The sampler includes a replay buffer that stores negative-energy samples from previous training iterations and uses them as initial points for later Langevin chains. 

\subsubsection*{Bayesian normalized autoencoder}

If we relate the NAE score $\etheta(x)$ to a likelihood, a statistical analysis also needs to control its precision. For sparse data, it can be unreliable and hence not directly usable. To capture this statistical uncertainty, we employ a Bayesian NAE (BNAE), replacing the deterministic network parameters with generalized parameter distributions~\cite{MacKay:1992bay, Neal:1995phd, Gal:2016phd, Haussmann:2021phd},
\begin{align}
    q_{\phi}(\theta) 
    \approx p(\theta\vert\dtrain) 
    = \frac{p(\dtrain\vert\theta)p(\theta)}{p(\dtrain)} \eqc
\end{align}
with a suitable prior $p(\theta)$, the training-irrelevant evidence $p(\dtrain)$, and the generic loss
\begin{align}
    \kl{q_{\phi}(\theta)}{p(\theta\vert\dtrain)} 
    &= - \Langle \log p(\dtrain\vert\theta) \Rangle_{q_{\phi}}
    + \kl{q_{\phi}(\theta)}{p(\theta)}
    + \text{const} \eqp
\end{align}
We apply this construction to the NAE loss in Eq.\eqref{eq:training_ebm} and get
\begin{align}
    \loss_{\text{BNAE}}(\phi)
    = \frac{1}{T} \XLangle \Langle \etheta(x) \Rangle_{\pdata} - \Langle \etheta(x) \Rangle_{\ptheta} \XRangle_{q_{\phi}} + \kl{q_{\phi}(\theta)}{p(\theta)} \eqp
    \label{eq:bnae_loss}
\end{align}
Choosing the weight distributions and the prior to be Gaussian, we can compute the KL-divergence analytically. The expectation value and variance of the learned energy,
\begin{align}
    \XXLangle \frac{\etheta(x)}{T} \XXRangle_{q_{\phi}} 
    &= \frac{1}{T} \int\dd\theta\: q_{\phi}(\theta) \etheta(x) \notag \\
    \var_{q_{\phi}}\left[\frac{\etheta(x)}{T}\right]
    &= \frac{1}{T^{2}} \int\dd\theta\: q_{\phi}(\theta) \left[ \etheta(x) - \mean{\etheta(x)}_{q_{\phi}} \right]^{2} \eqc
\end{align}
can be evaluated using Monte Carlo samples $\theta \sim q_{\phi}(\theta)$. This gives us the learned likelihood
\begin{align}
    \Langle \log\ptheta(x) \Rangle_{q_{\phi}}
    &= \XXLangle -\frac{\etheta(x)}{T}-\log\ztheta \XXRangle_{q_{\phi}} \notag \\
    \var_{q_{\phi}}\bigl[\log\ptheta(x)\bigr]
    &= \var_{q_{\phi}}\left[-\frac{\etheta(x)}{T}-\log\ztheta\right] \eqp
\end{align}

\clearpage
\section{Toy model}
\label{sec:toy}

We establish the relation of an uncertainty-aware NAE anomaly score with the underlying likelihood for a two-dimensional toy model where we know the true density
\begin{align}
    p_{2\gauss}(x)
    = \frac{1}{2} \gauss(x\vert \mu_1, \sigma^2 \mathbb{1})
    + \frac{1}{2} \gauss(x\vert \mu_2, \sigma^2 \mathbb{1}) \eqc
    \label{eq:true_prob_toy}
\end{align}
with 
\begin{align}
    \mu_1 =
    \begin{pmatrix}
        1.5 \\ 1.5
    \end{pmatrix}
    \qqqquad
    \mu_2 =
    \begin{pmatrix}
        -1.5 \\ -1.5
    \end{pmatrix} 
    \qqqquad 
    \sigma^2 = 0.5 \eqp
\end{align}
We compare the learned $\ptheta(x)$ to the true density across phase space. 

The MLPs for the encoder and decoder each consist of three hidden layers with $128$ neurons per layer. All layers use ReLU activations, except for the encoder and decoder outputs, which use linear activations. We use a three-dimensional Euclidean bottleneck with $d_{z} > d_{x}$, where the NAE regularization, described in Appendix~\ref{app:network_training}, prevents a trivial identity mapping~\cite{yoonAutoencodingNormalizationConstraints2023}. For the BNAE only the final decoder layer is Bayesian, as more Bayesian weights destabilize the training. 

We train the networks on $500$k points drawn from $\ptrue(x)$, for $100$ epochs using the \adam optimizer~\cite{Kingma:2014vow} with default parameters, a constant learning rate of $10^{-5}$, and a batch size of $2048$. For the toy model, the temperature is learnable, initialized at $T = 0.2$. We train it using the \adam optimizer with default parameters and a constant learning rate of $10^{-3}$. To stabilize the BNAE training, we first pre-train a Bayesian AE (BAE) for $100$ epochs at a constant learning rate of $10^{-3}$. We give more details and hyperparameters in Appendix~\ref{app:hyperparameters}.

For the evaluation, we use one test dataset $\dtest$ with $250$k points sampled from $\ptrue(x)$ and a test grid $\dgrid$, containing $500 \times 500$ points uniformly covering the region $[-4,4] \times [-4,4]$. To evaluate the BNAE, we compute the mean and variance of the learned density over $100$ Monte Carlo samples.

\subsection{Anomaly score metrics}
\label{sec:toy_metrics}

We apply four statistical tests to assess the accuracy and precision of the NAE anomaly score. 

\subsubsection*{Density accuracy}

An NAE learns the phase space density via the energy
\begin{align}
    \frac{\etheta(x)}{\ttheta} + \log\ztheta
    = -\log\ptheta(x)
    \approx -\log\ptrue(x) \eqp
\end{align}
We compute $\ztheta$ by numerically integrating $\exp(-\etheta(x)/\ttheta)$ over the test grid. We then compare the learned and true log-densities at each point $x$ and quantify their agreement using
\begin{alignat}{9}
    \Delta(x) &=
    \frac{\log\ptheta(x) -\log\ptrue(x)}{\log\ptrue(x)} 
    &&\qqquad \text{(NAE)} \notag \\
    \Delta(x) &=
    \frac{\mean{\log\ptheta(x)}-\log\ptrue(x)}{\log\ptrue(x)} 
    &&\qqquad \text{(BNAE)}\eqp
    \label{eq:rel_acc2}
\end{alignat}

\subsubsection*{Uncertainty pull}

To assess the calibration of the learned statistical uncertainties, we evaluate the local pull
\begin{align}
    t(x) = \frac{\mean{\log\ptheta(x)}-\log\ptrue(x)}{\sigstat(\log\ptheta)}
    \label{eq:pull}
\end{align}
on $\dtest$ with $\sigstat(\log\ptheta)$ as the standard deviation over Monte Carlo samples. We can define the pull for $\ptheta(x)$ by replacing $\log\ptheta(x)$ with $\ptheta(x)$ throughout.

\subsubsection*{Uncertainty calibration}

We test the calibration for Gaussian and non-Gaussian uncertainties~\cite{Bahl:2024gyt, Bahl:2025xvx}. For the BNAE, we construct uncertainty intervals directly from the Monte Carlo weight samples $\{\log\ptheta(x)\}$. Without assuming a distribution, the central $(1-\alpha)$-interval is defined by the empirical quantiles
\begin{align}
    C_{\alpha}(x)
    = \left[
        q_{\alpha/2}(x),
        q_{1-\alpha/2}(x)
    \right] \eqc
    \label{eq:quantiles}
\end{align}
where $q_{\tau}(x)$ is the empirical $\tau$-quantile of $\{\log\ptheta(x)\}$ at fixed $x$. For Gaussian uncertainties, we can construct $C_{\alpha}(x)$ from the $p$-value of $\log\ptrue(x)$ with mean $\mean{\log\ptheta(x)}$ and standard deviation $\sigstat(\log\ptheta)$,
\begin{align}
    p_{\text{Gauss}}
    = 2\left[1-\Phi\left(\frac{\vert \log\ptrue(x)-\mean{\log\ptheta(x)} \vert}{\sigstat(\log\ptheta)}\right)\right] \eqp
\end{align}
Here, $\Phi$ is the standard normal cumulative distribution function. The confidence interval $C_{\alpha}(x)$ at nominal level $(1-\alpha)$ is defined by all log-density values for which the $p$-value exceeds $\alpha$, 
\begin{align}
    C_{\alpha}(x) = \Big[
        \mean{\log\ptheta(x)} &- \Phi^{-1}(1-\alpha/2)\sigstat(\log\ptheta), \notag \\
        \mean{\log\ptheta(x)} &+ \Phi^{-1}(1-\alpha/2)\sigstat(\log\ptheta) \Big] \eqp
    \label{eq:gaussian}
\end{align} 

For a nominal confidence level $(1-\alpha)$, the marginal coverage is the probability that the true log-density lies inside the predicted interval,
\begin{align}
    c_{1-\alpha}^{\text{marg}} =
    \int \dd x\: p(x)\;
    \mathbb{1}\!\left\{
        \log\ptrue(x) \in C_{\alpha}(x)
    \right\} \eqc
    \label{eq:marg_cov}
\end{align}
where $p(x)$ denotes the phase-space probability. We evaluate this integral on $\dtest$. Calibrated intervals satisfy
\begin{align}
    c_{1-\alpha}^{\text{marg}} = 1-\alpha
    \qquad\text{for all $\alpha$} \eqp
\end{align}
If $c_{1-\alpha}^{\text{marg}} > 1-\alpha$, the uncertainties are overestimated or conservative; if $c_{1-\alpha}^{\text{marg}} < 1-\alpha$, the uncertainties are underestimated. All definitions above apply analogously to $\ptheta(x)$, replacing $\log\ptheta(x)$ by $\ptheta(x)$ throughout.

\subsubsection*{Classifier test}

Finally, we train a classifier to distinguish samples from $\ptrue(x)$ and samples from $\ptheta(x)$. We generate $250$k (B)NAE points using LMC, and $250$k reference points from $\ptrue(x)$. The combined dataset is split into $300$k training, $50$k validation, and $150$k test points. We use an MLP with three hidden layers of $64$ neurons each. The hidden layers use ReLU activations, the output layer uses a sigmoid. We train the classifier on the binary cross-entropy with the \adam optimizer, using default parameters, a constant learning rate of $10^{-3}$, and a batch size of $2048$. We train for at most $200$ epochs, with early stopping based on the validation loss and a patience of $20$ epochs.

\subsection{NAE results}
\label{sec:toy_nae}

\begin{figure}[t]
    \includegraphics[page=01, width=0.45\linewidth]{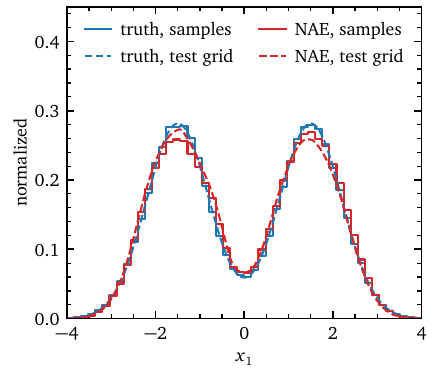}\hfill
    \includegraphics[page=02, width=0.45\linewidth]{figures/toy_2gauss_nae.pdf}
    \caption{Marginalized true and learned NAE densities of the two-dimensional toy model, evaluated on the test grid. The classifier test yields an AUC of $0.536$.}
    \label{fig:toy_nae_marginal}
\end{figure}

\begin{figure}[b!]
    \includegraphics[page=05, width=0.45\linewidth]{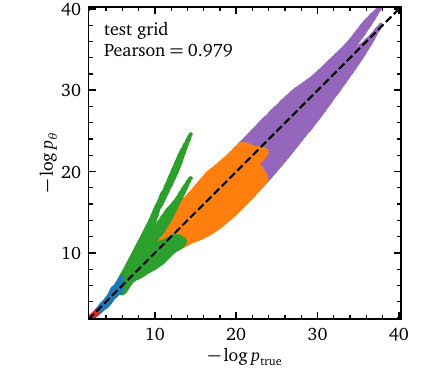}\hfill
    \includegraphics[page=06, width=0.45\linewidth]{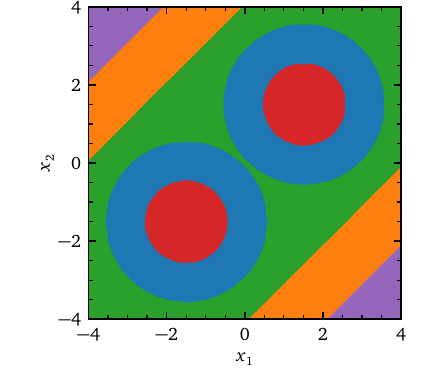}
    \caption{Learned versus true log-likelihoods for the NAE (left) and corresponding areas on the test grid (right).}
    \label{fig:toy_nae_regions1}
\end{figure}

\begin{figure}[t]
    \includegraphics[page=03, width=0.45\linewidth]{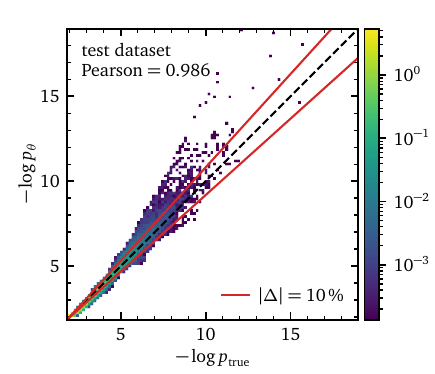}\hfill
    \includegraphics[page=04, width=0.45\linewidth]{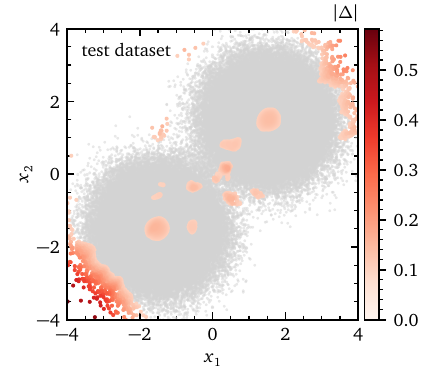}
    \caption{Learned versus true log-likelihoods for the NAE (left) and points with $\vert\Delta(x)\vert \geq 10\,\%$ (right) on $\dtest$. The light gray points have $\vert\Delta(x)\vert < 10\,\%$.}
    \label{fig:toy_nae_regions2}
\end{figure}

In \cref{fig:toy_nae_marginal} we compare the marginalized $\ptrue(x)$ and $\ptheta(x)$ on our test grid. The NAE captures the structure of the target distribution, with a consistent slight underestimate on the Gaussian peaks. 

To understand the source of the small discrepancies, we investigate the correlation between $\log\ptheta(x)$ and $\log\ptrue(x)$ on the same test grid in \cref{fig:toy_nae_regions1}. In the left panel, we see the diagonal correlation. It indicates that the density, learned via the energy, tracks the true density over the full range of likelihood values, with a Pearson correlation coefficient of $0.979$. To identify where deviations from $\ptrue(x)$ arise, we show representative $\dgrid$ regions in both panels. For the Gaussian peak regions the learned and true likelihoods agree well, whereas points farther away from the peaks exhibit increasingly large deviations. 

In \cref{fig:toy_nae_regions2}, we evaluate the learned density on $\dtest$. In the left panel, we show the likelihood correlation, with the red lines depicting the region where $\vert\Delta(x)\vert = 10\,\%$. In the right panel we see that most phase-space points satisfy $\vert\Delta(x)\vert < 10\,\%$. We highlight points with $\vert\Delta(x)\vert \geq 10\,\%$, predominantly in the tails and near the peaks of the distributions. The light gray points have $\vert\Delta(x)\vert < 10\,\%$.

A supervised classifier trained to distinguish samples drawn from $\ptheta(x)$ and $\ptrue(x)$ gives
\begin{align} 
    \text{AUC}\big\vert_\text{NAE} = 0.536 \eqc
\end{align}
providing further evidence that the NAE indeed learns the target density.

\subsection{BNAE results}
\label{sec:toy_bnae}

\begin{figure}[b!]
    \includegraphics[page=01, width=0.325\linewidth]{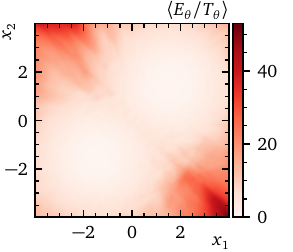}\hfill
    \includegraphics[page=02, width=0.325\linewidth]{figures/toy_2gauss_bnae.pdf}\hfill
    \includegraphics[page=03, width=0.325\linewidth]{figures/toy_2gauss_bnae.pdf}\\
    \includegraphics[page=04, width=0.325\linewidth]{figures/toy_2gauss_bnae.pdf}\hfill
    \includegraphics[page=05, width=0.325\linewidth]{figures/toy_2gauss_bnae.pdf}\hfill
    \includegraphics[page=06, width=0.325\linewidth]{figures/toy_2gauss_bnae.pdf}
    \caption{Learned BNAE energy, energy uncertainty, and relative energy uncertainty of the two-dimensional toy model, evaluated on the test grid (upper panels). Below, we show the same plots for the learned density.}
    \label{fig:toy_bnae_density}
\end{figure}

The BNAE learns the target density with meaningful statistical uncertainties. In the upper panels of \cref{fig:toy_bnae_density} we show the learned temperature-rescaled energy and its uncertainty evaluated on $\dgrid$. Both are small near the data support and increase in the low-density regions. The relative energy uncertainty peaks in the high-density regions, where the denominator $\mean{\etheta(x)/\ttheta}$ becomes small. 

In the lower panels of \cref{fig:toy_bnae_density} we show the corresponding density and its uncertainty evaluated on $\dgrid$. The mean density follows the Gaussians whereas its uncertainty peaks in a ring around them. Here, the density changes rapidly. We find that the relative density uncertainty is smaller in the high-density regions and larger in the low-density regions. Ignoring the uncertainty in $\ztheta$, density fluctuations are driven by energy fluctuations, so that 
\begin{align}
    \frac{\sigstat(\ptheta)}{\mean{\ptheta(x)}} 
    \approx \sigstat\left(\frac{\etheta(x)}{\ttheta}\right) \eqp
\end{align}
Indeed, we observe in \cref{fig:toy_bnae_density} that the absolute energy uncertainty and the relative density uncertainty behave the same. 

\begin{figure}[t]
    \includegraphics[page=07, width=0.45\linewidth]{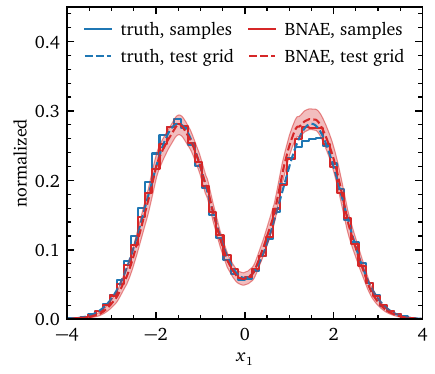}\hfill
    \includegraphics[page=08, width=0.45\linewidth]{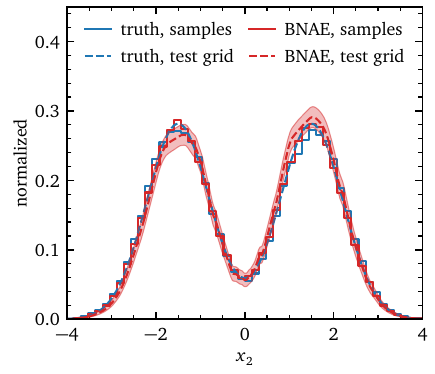}
    \caption{Marginalized true and learned BNAE densities of the two-dimensional toy model, evaluated on the test grid. We show $\pm 1\sigma$ uncertainty bands from $100$ Monte Carlo samples. The classifier test yields an AUC of $0.522$.}
    \label{fig:toy_bnae_marginal}
\end{figure}

\begin{figure}[b!]
    \includegraphics[page=13, width=0.45\linewidth]{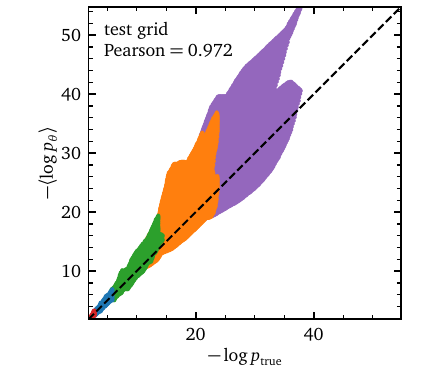}\hfill
    \includegraphics[page=14, width=0.45\linewidth]{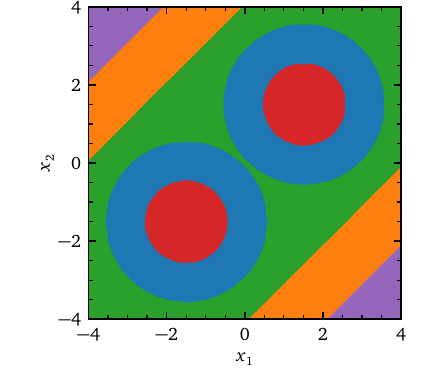}
    \caption{Learned versus true log-likelihoods for the BNAE (left) and corresponding areas on the test grid (right).}
    \label{fig:toy_bnae_regions1}
\end{figure}

In \cref{fig:toy_bnae_marginal} we show the marginal distributions of $\mean{\ptheta(x)}$ and their uncertainties. The qualitative behavior of the marginal densities is the same as that of the NAE shown in \cref{fig:toy_nae_marginal}. Analogously to \cref{fig:toy_nae_regions1}, we show the BNAE mapping in \cref{fig:toy_bnae_regions1}. The main difference from the NAE is the larger deviations in the off-diagonal corners, far away from the Gaussian peaks. For $-\log\ptrue(x) > 40$ the learned likelihood no longer agrees with the truth.

\begin{figure}[t]
    \includegraphics[page=09, width=0.45\linewidth]{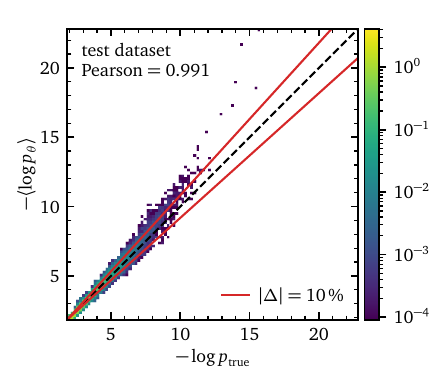}\hfill
    \includegraphics[page=10, width=0.45\linewidth]{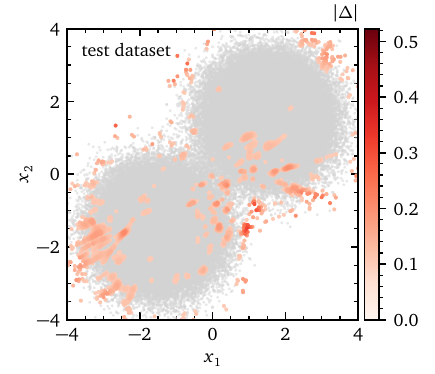}
    \caption{Learned versus true log-likelihoods for the BNAE (left) and points with $\vert\Delta(x)\vert \geq 10\,\%$ (right) on $\dtest$. The light gray points have $\vert\Delta(x)\vert < 10\,\%$.}
    \label{fig:toy_bnae_regions2}
\end{figure}

In \cref{fig:toy_bnae_regions2} we show the corresponding likelihood-correlation plot on $\dtest$ instead. On the actual test data the learned likelihood reproduces the truth better than for the NAE in \cref{fig:toy_nae_regions2}. Again, we divide the test data points along $\vert\Delta(x)\vert$. Altogether, the BNAE learns $\mean{\log\ptheta(x)}$ accurately, with the largest deviations in the tails of the distribution. 

If we compare \cref{fig:toy_nae_regions1,fig:toy_bnae_regions1}, we see that the NAE and the BNAE extrapolate into the tails of the distribution differently (orange and purple regions). On the test grid, the NAE in \cref{fig:toy_nae_regions1} captures the tails of the distribution well, while the BNAE in \cref{fig:toy_bnae_regions1} increasingly deviates from the diagonal toward the outer regions. The BNAE may compensate for the lack of training data in these regions by increasing its statistical uncertainty, which can come at the expense of an accurate mean prediction. However, this extrapolation behavior varies between different network trainings.

The BNAE classifier test gives 
\begin{align}
    \text{AUC}\big\vert_\text{BNAE} = 0.522 \eqc
\end{align}
so the samples generated by the BNAE closely match the true density. Again, although we observe an improvement in the AUC from the NAE to the BNAE, these numbers vary between network trainings.

\begin{figure}[b!]
    \includegraphics[page=11, width=0.45\linewidth]{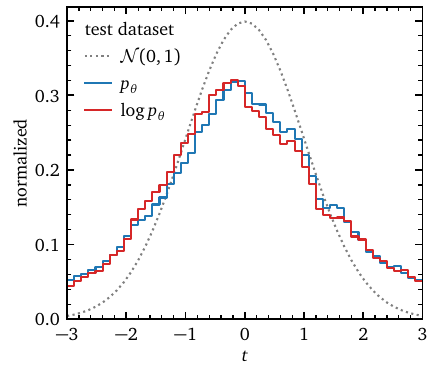}\hfill
    \includegraphics[page=12, width=0.45\linewidth]{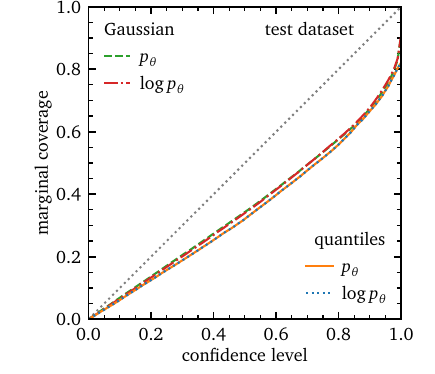}
    \caption{Pulls (left) and marginal coverages (right) of the two-dimensional toy model, evaluated on the test dataset. For the coverages, we show the quantile-based construction of Eq.\eqref{eq:quantiles} and the Gaussian intervals from Eq.\eqref{eq:gaussian}.}
    \label{fig:toy_bnae_coverage}
\end{figure}

In the left panel of \cref{fig:toy_bnae_coverage}, we show the statistical pulls of $\ptheta(x)$ and $\log\ptheta(x)$, defined in Eq.\eqref{eq:pull}, evaluated on $\dtest$. Both pulls are approximately Gaussian. In the right panel we show the marginal coverage, defined in Eq.\eqref{eq:marg_cov}, for $\ptheta(x)$ and $\log\ptheta(x)$, evaluated on $\dtest$ using the quantile-based intervals of Eq.\eqref{eq:quantiles} and the Gaussian intervals of Eq.\eqref{eq:gaussian}. We observe a small under-coverage, \ie, the learned uncertainties are slightly too small. 

The quantile-based coverage curves for $\ptheta(x)$ and $\log\ptheta(x)$ are indistinguishable. This is expected, as the empirical quantiles are computed from the same underlying Monte Carlo samples and the logarithm is monotonic, so the resulting coverage is identical by construction. The same does not hold for the Gaussian construction, since a Gaussian in $\ptheta(x)$ does not map to a Gaussian in $\log\ptheta(x)$. However, comparing the Gaussian-based and quantile-based coverages, we see that for $\ptheta(x)$ and for $\log\ptheta(x)$, the curves agree closely. In Appendix~\ref{app:results}, the marginal coverages for the colored regions in \cref{fig:toy_bnae_regions1} show that the learned uncertainties are best calibrated in the region around the two Gaussian peaks and are underestimated as we move towards the corners.

\subsection{Dual-NAE test}
\label{sec:toy_lr_test}

\begin{figure}[b!]
    \includegraphics[page=02, width=0.45\linewidth]{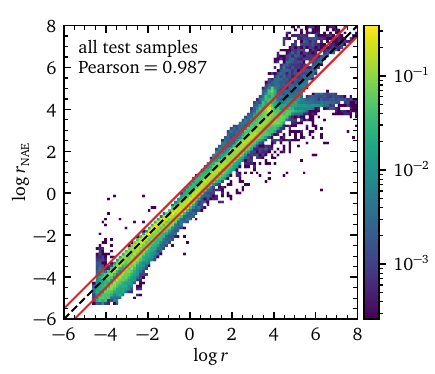}\hfill
    \includegraphics[page=03, width=0.45\linewidth]{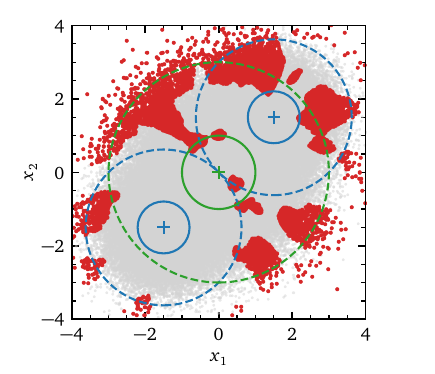}
	\caption{Left: NAE-extracted versus true log-likelihood ratios of the toy model, evaluated on test samples drawn from $p_{1\gauss}(x)$ and $p_{2\gauss}(x)$. Right: test points outside the boundaries in red, points inside the boundaries in light gray.}
    \label{fig:toy_likelihood_test1}
\end{figure}

When we do not know the true density, we can use two learned densities to construct a likelihood ratio. Training a classifier on samples from these two densities also provides an approximation to the likelihood ratio. A promising avenue then is to train two NAEs on two different densities and correlate their learned energies to the supervised classifier. We complement our toy model from Eq.\eqref{eq:true_prob_toy} with a second density
\begin{align}
    p_{1\gauss}(x) = \gauss(x\vert 0, \mathbb{1}) \eqc
\end{align}
such that a supervised classifier would learn
\begin{align}
    r(x) = \frac{p_{1\gauss}(x)}{p_{2\gauss}(x)}
    = \frac{2\gauss(x\vert 0,\mathbb{1})}{\gauss(x\vert \mu_1, \sigma^2 \mathbb{1})
    + \gauss(x\vert \mu_2, \sigma^2 \mathbb{1})} \eqp
    \label{eq:true_llr}
\end{align}
We then train one NAE on samples drawn from $p_{1\gauss}(x)$ and one NAE on samples from $p_{2\gauss}(x)$. This gives us
\begin{align}
    p_{\theta,i}(x) =
    \frac{e^{-E_{\theta,i}(x)/T_{\theta,i}}}{Z_{\theta,i}}
    \qquad\mwith\quad
    i \in \{ 1\gauss, 2\gauss \} \eqc
\end{align}
where $E_{\theta,i}(x)$ are the learned energies, $T_{\theta,i}$ the learned temperatures, and $Z_{\theta,i}$ the partition functions. The NAE-extracted log-likelihood ratio $r_\text{NAE}$ is defined as
\begin{align}
    \log r_{\text{NAE}}(x)
    = \frac{E_{\theta,2\gauss}(x)}{T_{\theta,2\gauss}}
    - \frac{E_{\theta,1\gauss}(x)}{T_{\theta,1\gauss}}
    + \ctheta
    \qquad\mwith\quad
    \ctheta = -\log\frac{Z_{\theta,1\gauss}}{Z_{\theta,2\gauss}} \eqp
    \label{eq:loglr_toy}
\end{align}
In \cref{fig:toy_likelihood_test1}, we compare the true and NAE-extracted log-likelihood ratios on test samples drawn from $p_{1\gauss}(x)$ and $p_{2\gauss}(x)$, setting $\ctheta=0$. In the left panel we show the correlation between the NAE-extracted log-likelihood ratio and the true log-likelihood ratio, defined in Eq.\eqref{eq:true_llr}. The red boundaries mark regions with large deviations. We highlight the corresponding phase-space points in red in the right panel. The green contours represent $p_{1\gauss}(x)$ and the blue contours $p_{2\gauss}(x)$ at the $1\sigma$ (solid) and $3\sigma$ (dashed) levels. The large-deviation points lie in regions where at least one density lacks statistical support. Since $E_{\theta,1\gauss}(x)$ and $E_{\theta,2\gauss}(x)$ are trained separately on the two distributions, either can trigger a breakdown. The red points in the right panel are scattered in the tails of $p_{1\gauss}(x)$ and in the inter-mode region and outer tails of $p_{2\gauss}(x)$. 

\begin{figure}[t]
    \centering
    \includegraphics[page=01, width=0.45\linewidth]{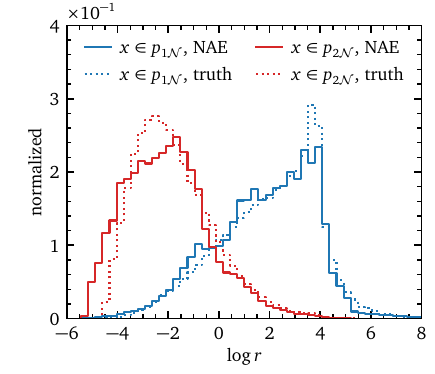}
	\caption{NAE-extracted versus true log-likelihood ratios of the toy model, evaluated on each of the two test samples from $p_{1\gauss}(x)$ and $p_{2\gauss}(x)$.}
    \label{fig:toy_likelihood_test2}
\end{figure}

In \cref{fig:toy_likelihood_test2} we compare the NAE-extracted log-likelihood ratios with their truth counterparts for the two reference samples separately. The temperatures are learned, and we set $\ctheta=0$ even though a nonzero $\ctheta$ would allow us to shift the two distributions jointly. Following the definition of Eq.\eqref{eq:true_llr}, the points from $p_{1\gauss}(x)$ dominate at large $r(x)$, whereas points from $p_{2\gauss}(x)$ cluster at small log-likelihood ratios. The combination indicates that each NAE learns the correct likelihood for the respective training sample, such that their combination predicts the correct likelihood ratio. Unlike the results from \cref{sec:toy_nae,sec:toy_bnae}, this test can be applied to cases where the individual likelihoods are not known, but we can train a supervised classifier as a likelihood ratio reference.

\clearpage
\section{Jet tagging}
\label{sec:jets}

As a proper LHC benchmark we use the public \toptagging dataset~\cite{Kasieczka:2019dbj}, which contains top jet and QCD jet events generated with \pythia~8.2.15~\cite{Sjostrand:2014zea} at $14\,\tev$ without multi-parton interactions and pile-up. Detector effects are simulated with \delphes~3.3.2~\cite{deFavereau:2013fsa} using the default ATLAS card and we cluster the jets using the anti-$\kt$ algorithm~\cite{Cacciari:2008gp} in \fastjet~3.1.3~\cite{Cacciari:2011ma} with $R=0.8$. We only retain the leading jet in each event, requiring
\begin{align}
    \ptjet \in [550,650]\,\gev
    \qquad\mand\qquad 
    \vert\etajet\vert < 2.0 \eqp
\end{align}    
The dataset consists of $2$M jets, equally split into QCD and top jets. It contains only kinematic information, no tracking or particle identification information. As outlined in Appendix~\ref{app:jets_details} we preprocess the jets into jet images and apply a Gaussian filter to reduce the sparsity. 

For the (B)NAE, we use a convolutional autoencoder with a six-dimensional latent space, summarized in \cref{tab:architecture} in Appendix~\ref{app:hyperparameters}. We either train the (B)NAE on QCD jets and test it on anomalous top jets (top-tagging), or we train the (B)NAE on top jets and test it on anomalous QCD jets (QCD-tagging). Each training direction uses $100$k training jets, $50$k validation jets, and $200$k test jets.

To initialize the parameters and the decoder manifold, we pre-train a standard (B)AE for $500$ epochs using \adam with default parameters, a learning rate of $10^{-3}$, and a batch size of $2048$. Then, we train the (B)NAE for $100$ epochs with a learning rate of $10^{-5}$, and a batch size of $128$. To evaluate the BNAE, we compute the mean and variance of the learned energy and the corresponding ROC curves over $50$ Monte Carlo samples. We provide further details in Appendix~\ref{app:hyperparameters}.

For the dual-(B)NAE test with top and QCD jets, we train a supervised classifier with the same architecture as the NAE encoder in \cref{tab:architecture}. The hidden layers use parametric ReLU (PReLU) activations, while the output layer uses a sigmoid. For training, we use \adam with a learning rate of $10^{-3}$ and a batch size of $2048$. The training, validation, and test datasets contain $200$k, $100$k, and $200$k jets. We train for $300$ epochs and then use the network with the lowest validation loss.

\subsection{Performance with uncertainties}
\label{sec:jets_performance}

\begin{figure}[b!]
    \includegraphics[page=01, width=0.45\linewidth]{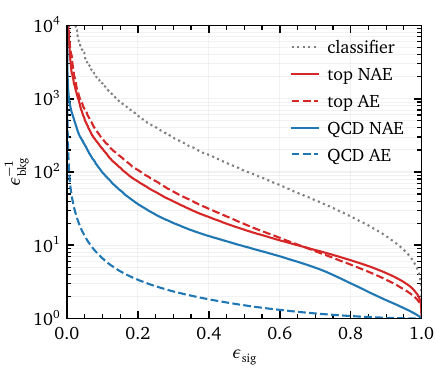}\hfill
    \includegraphics[page=01, width=0.45\linewidth]{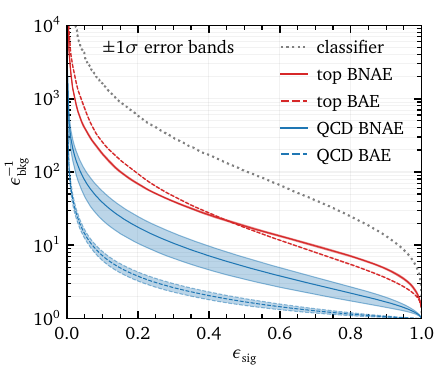}
    \caption{ROC curves for (B)NAE top-tagging and QCD-tagging. For the BNAE in the right panel we show $\pm 1\sigma$ uncertainty bands from $50$ Monte Carlo samples. We show a supervised classifier benchmark as a dashed line.}
    \label{fig:nae-tt-performance}
\end{figure}

We evaluate the NAE and BNAE as unsupervised anomaly detectors for top-tagging and QCD-tagging. In each case, we train only on background jets and use the learned energy as the anomaly score. In \cref{fig:nae-tt-performance}, we show the ROC curves for top-tagging (red) and QCD-tagging (blue), with the AE/NAE results in the left panel and the BAE/BNAE results in the right panel. We summarize the AUC values and background rejections in \cref{tab:nae-tt-performance}.

For the deterministic networks, we reproduce the qualitative findings of Ref.~\cite{Dillon:2022mkq}: the AE performs worse for QCD-tagging than for top-tagging, while the NAE largely removes this asymmetry. The ROC curves in \cref{fig:nae-tt-performance} (left) clearly show this. For QCD-tagging, our AE with a six-dimensional latent space achieves an AUC of $0.39$, lower than that of the three-dimensional latent space AE from Ref.~\cite{Dillon:2022mkq}. We obtain an AUC below $0.50$ because the AE trained on top jets reconstructs the less complex QCD jets well and therefore assigns them lower anomaly scores. Our QCD-tagging NAE performs worse than the NAE in Ref.~\cite{Dillon:2022mkq}. However, Ref.~\cite{Dillon:2022mkq} does not consider the learned underlying density, whereas we select hyperparameters so that our energy-based networks correctly learn the training data density as we will show in \cref{sec:jets_lr_test}.

In \cref{tab:nae-tt-performance} we see that the BAE shows the same tagging asymmetry as the AE and achieves very similar performance within learned uncertainties. The BNAE performance is compatible with the NAE baseline, but for QCD-tagging barely and with a tendency of a reduced performance. The error bars in \cref{fig:nae-tt-performance} (right) and \cref{tab:nae-tt-performance} show that both the BAE and BNAE produce larger uncertainties for QCD-tagging than for top-tagging. For QCD-tagging, the AUC uncertainties reach order $10^{-2}$, compared to $10^{-3}$ for top-tagging. Even after accounting for these large uncertainties, the BNAE still outperforms the BAE for QCD-tagging significantly. In \cref{fig:jets_bnae_energies} of Appendix~\ref{app:bnae_energies}, we show that the BNAE learns physically reasonable energy distributions: it assigns lower energies to the background samples we use for training and higher energies to samples outside the training distribution.

\begin{table}[t]
    \centering
    \begin{small}
        \begin{tabular}{@{\hspace{2pt}}ll|S[table-format=1.4(1.4)]S[table-format=3.1(1.1)]S[table-format=3.1(1.1)]@{\hspace{2pt}}}
            \toprule
            task & network & {$\rocauc$} & {$\bkgrej(\sigeff=0.2)$} & {$\bkgrej(\sigeff=0.3)$} \\
            \midrule
            \multicolumn{2}{@{\hspace{2pt}}l|}{supervised classifier} & 0.9728 & 584.8 & 294.6 \\
            \midrule
            top-tagging & AE   & 0.8947         & 106.9     & 54.2 \\
            top-tagging & NAE  & 0.9036         & 74.5      & 39.8 \\
            QCD-tagging & AE   & 0.3912         & 3.4       & 2.3 \\
            QCD-tagging & NAE  & 0.8071         & 36.5      & 20.1 \\
            \midrule
            top-tagging & BAE  & 0.8959(0.0008) & 94.1(0.8) & 46.8(0.4) \\
            top-tagging & BNAE & 0.9126(0.0020) & 69.0(1.4) & 39.7(0.6) \\
            QCD-tagging & BAE  & 0.4319(0.0368) & 3.8(0.5)  & 2.7(0.4) \\
            QCD-tagging & BNAE & 0.7292(0.0536) & 17.7(7.0) & 10.7(4.0) \\
            \bottomrule
        \end{tabular}
    \end{small}
    \caption{Top-tagging and QCD-tagging performance after (B)AE pre-training and after (B)NAE training. We report $\pm 1\sigma$ uncertainties from $50$ Monte Carlo samples for the Bayesian networks. The supervised classifier provides a benchmark.}
    \label{tab:nae-tt-performance}
\end{table}

\subsection{Statistical interpretation}
\label{sec:jets_lr_test}

\begin{figure}[t]
	\centering
    \includegraphics[page=01, width=0.45\linewidth]{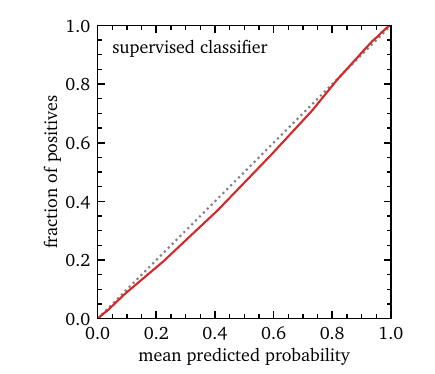}
	\caption{Calibration of the supervised classifier, showing the observed fraction of top jets in the test dataset as a function of the mean predicted probability.}
    \label{fig:clf_reliability}
\end{figure}

To test whether the (B)NAEs learn the underlying top-jet and QCD-jet densities, we compare their temperature-rescaled energy difference to the likelihood ratio inferred from a supervised classifier, following the methodology of \cref{sec:toy_lr_test}. Unlike for the toy model, we do not have access to the true likelihood ratio, so we train a supervised classifier to learn the likelihood ratio,
\begin{align}
    r_{\text{class}}(x)
    \equiv \frac{\class(x)}{1-\class(x)}
    \approx \frac{p_{\text{top}}(x)}{p_{\text{QCD}}(x)} \eqp
    \label{eq:lr_clf}
\end{align}
In \cref{fig:clf_reliability} we show the observed fraction of top jets as a function of the classifier score $\class(x)$. We see that the classifier is well calibrated and we can use it to approximate the true likelihood ratio.

\subsubsection*{Dual-NAE test}

\begin{figure}[t]
    \includegraphics[page=02, width=0.45\linewidth]{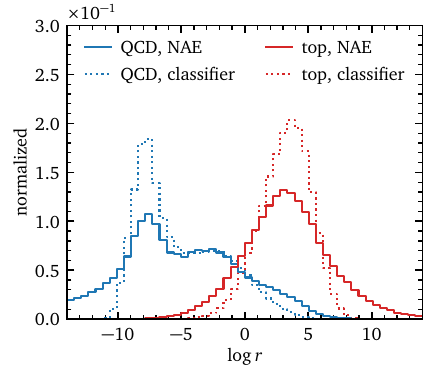}\hfill
    \includegraphics[page=03, width=0.45\linewidth]{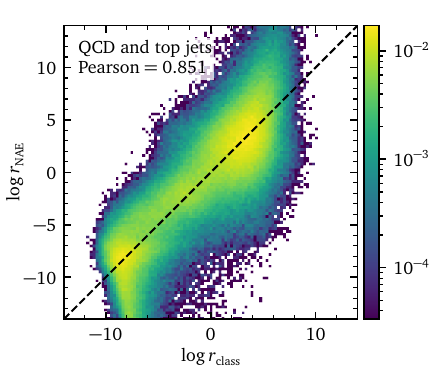}\\
    \includegraphics[page=02, width=0.45\linewidth]{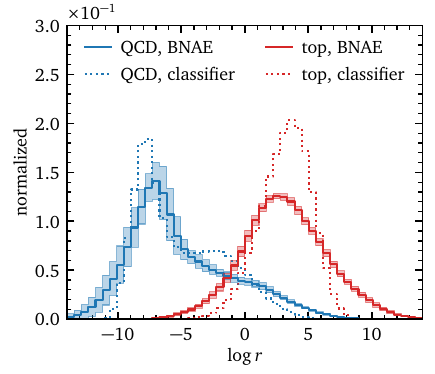}\hfill
    \includegraphics[page=03, width=0.45\linewidth]{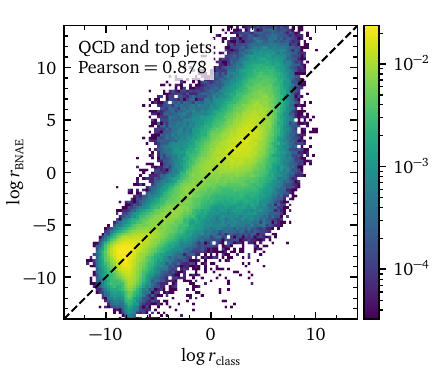}
	\caption{(B)NAE-extracted versus classifier-based log-likelihood ratio for jet tagging. We obtain the $\pm 1\sigma$ uncertainty bands by evaluating the log-likelihood ratio for all pairwise combinations of Monte Carlo samples from the two separately trained BNAEs.}
	\label{fig:llr-tt-check}
\end{figure}

If the two (B)NAEs learn the correct densities, the difference between the temperature-rescaled energies should reproduce $\log r_{\text{class}}(x)$ up to a constant. Similar to  \cref{fig:toy_likelihood_test1,fig:toy_likelihood_test2}, we compare the classifier-based likelihood ratios with the (B)NAE-extracted likelihood ratios in \cref{fig:llr-tt-check}. The top panels show the NAE result, while the bottom panels show the BNAE result with uncertainties. Both networks reproduce the classifier structure, but underestimate the peaks. Their tails extend further than those of $\log r_{\text{class}}(x)$. 

The top-right panel shows the correlations between $\log r_{\text{NAE}}(x)$ and $\log r_{\text{class}}(x)$ for the NAE, and the bottom-right panel for the BNAE. The points lie predominantly near the diagonal. As in \cref{sec:toy_lr_test}, we expect deviations primarily in regions with limited support from one or both training distributions. Unlike in the two-dimensional toy model, these regions are difficult to visualize in the high-dimensional jet-image space.

To test whether the (B)NAE reproduces the likelihood ratio quantitatively we fit a linear dependence with a free slope and additive constant,
\begin{align}
    \log r_{\text{class}}
    = m \frac{E_{\theta,\text{QCD}}-E_{\theta,\text{top}}}{T} + \ctheta \eqc
    \label{eq:m_fit}
\end{align}
to the data in the right panels of \cref{fig:llr-tt-check}. If the temperature-rescaled energy difference reproduces the true log-likelihood ratio up to the additive constant $\ctheta$, we expect $m = 1$. 

Unlike for the toy model in \cref{sec:toy_lr_test}, where a learned temperature $\ttheta$ and $\ctheta = 0$ give near-perfect agreement with a Pearson correlation coefficient of $0.987$, the jet-tagging case is less straightforward. Training is considerably less stable, and we find
\begin{align}
    \ctheta &= -1.9109 \qqqquad \text{(NAE)} \notag \\
    \ctheta &= -3.0292 \qqqquad \text{(BNAE)} \eqc
\end{align}
but $m \neq 1$ for many independent trainings with a learned temperature. Fixing $T$ gives us more control over the scale of the learned energy difference. We choose $T$ near the average MSE of the pre-trained (B)AE, which keeps $\etheta(x)/T \sim \order(1)$. Starting from there, training with smaller $T$ gives larger fitted slopes $m$, so we can select $T$ by requiring $m \approx 1$. As a test of this procedure we find Pearson correlation coefficients of $0.851$ for the NAE and $0.878$ for the BNAE, providing evidence for an approximate linear relation between the learned energy differences and $\log r_{\text{class}}(x)$. For practical applications, the dual-(B)NAE test allows us to adjust the temperature on a simulated background sample, combined with a hypothetical simulated signal sample. At this point we welcome ideas to learn or extract the correct temperature for high-dimensional feature spaces.

For the toy model, we showed in \cref{sec:toy_nae,sec:toy_bnae} that when the (B)NAE learns the density, it also generates negative-energy samples that closely match the target distribution. Jet images present a substantially more challenging problem because of their high dimensionality and sparsity. While the (B)NAE-generated jet images are easily separated from true jet images using a supervised classifier, they do reproduce several substructure observables, as shown in Appendix~\ref{app:nae_generated_samples}. 

\clearpage
\section{Outlook}
\label{sec:outlook}

Unsupervised anomaly detection with autoencoders provides an exciting model-agnostic approach to BSM searches at the LHC. However, the reconstruction error of a standard autoencoder defines an anomaly score without statistical interpretation. The normalized autoencoder trains the reconstruction error as the energy function of an energy-based network. In this work, we asked two questions: what statistical quantity does the NAE score represent, and can we extract it with a well-defined uncertainty? 

To the first question, we showed that the NAE score corresponds to the negative log-likelihood of a phase-space point because the energy-based training optimizes the likelihood. To the second question, we constructed a Bayesian NAE by making the final decoder layer Bayesian. It learned the density and its statistical uncertainty.

For a two-dimensional Gaussian-mixture toy model, the NAE reproduced the true log-likelihood within $10\,\%$ over most of the phase space, with a Pearson correlation of $0.979$. Classifier tests with the generated samples gave AUC values of $0.536$ (NAE) and $0.522$ (BNAE). The BNAE uncertainties increased in low-density regions, the pulls were approximately Gaussian, and the marginal coverage showed mild under-coverage.

For realistic LHC applications, the true likelihood is unknown, and we introduced a new dual-NAE test. It combines two (B)NAEs trained on complementary hypotheses to extract the likelihood ratio up to an additive constant. For the toy model, the NAEs reproduced the true likelihood ratio in regions with sufficient support. For jet tagging, we compared the temperature-rescaled energy difference with a classifier likelihood ratio and found Pearson correlations of $0.851$ for the NAE and $0.878$ for the BNAE. The generated jet images reproduced several high-level jet-substructure observables well.

\section*{Acknowledgements}

We would like to thank Barry Dillon, Luigi Favaro, and Sangwoong Yoon for their collaboration during a very early phase of this project, and Henning Bahl and Ayodele Ore for many inspiring discussions and for their continued support. This work is supported by the Deutsche Forschungsgemeinschaft (DFG, German Research Foundation) under grant 396021762 -- TRR 257 \textsl{Particle Physics Phenomenology after the Higgs Discovery}. RD acknowledges support from the Alexander von Humboldt Foundation.

\appendix

\clearpage
\section{Jet image preprocessing}
\label{app:jets_details}

A jet image is a two-dimensional representation of the energy distribution in the calorimeter. To standardize the inputs and improve training robustness, we center, rotate, flip, pixelize, crop, and normalize each jet. First, we center each jet by shifting the $\pt$-weighted centroid of its constituents to the origin in the \etaphi plane. We then rotate the jet so that its major principal axis points vertically toward increasing $\phi$ ($12$ o'clock) and flip it along both axes so that the highest-$\pt$ region lies in the upper-right quadrant. We pixelize the jet constituents in the \etaphi plane and crop the resulting image to $40\times40$ pixels, spanning $\eta=\range{-0.58}{0.58}$ and $\phi=\range{-0.70}{0.70}$ with pixel size $[\Delta\eta,\Delta\phi]=[0.029,0.035]$. We define each pixel intensity as the summed $\pt$ of the particle-flow objects within the corresponding $\eta$-$\phi$ cell and normalize the image by its total intensity to remove the overall $\pt$ scale. This prevents the network from learning trivial scale differences as anomalies. To reduce the sparsity, we apply a Gaussian filter of width $\sigma_{\text{GF}} = 1$ to each image.

\begin{figure}[H]
	\includegraphics[page=01, width=0.45\linewidth]{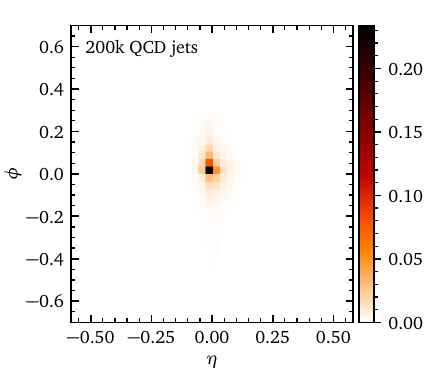}\hfill
	\includegraphics[page=02, width=0.45\linewidth]{figures/jets_test_data_images.pdf}\\
	\includegraphics[page=03, width=0.45\linewidth]{figures/jets_test_data_images.pdf}\hfill
	\includegraphics[page=04, width=0.45\linewidth]{figures/jets_test_data_images.pdf}
	\caption{Jet images from the \toptagging test dataset after preprocessing, for QCD jets (left) and top jets (right). We show the average over $200$k individual jet images (upper) and individual jets with a Gaussian filter applied (lower).}
	\label{fig:toptagging-images}
\end{figure}

In \cref{fig:toptagging-images}, we show preprocessed QCD (left) and top-jet (right) images from the \toptagging test dataset. In the upper panels we show the average pixel intensity over $200$k jets and in the lower panels we show individual jet images with a Gaussian filter applied. The images clearly reveal the characteristic one-prong structure of QCD jets and three-prong structure of top jets. Individual jet images are extremely sparse, with only $1\,\%$ to $2\,\%$ of the $1600$ pixels containing nonzero $\pt$.

\clearpage
\section{Hyperparameters}
\label{app:hyperparameters}

In this section, we provide additional details on the network architectures and hyperparameter settings used for the toy model in \cref{sec:toy} and for jet tagging in \cref{sec:jets}. We implement all networks using \pytorch~\cite{Paszke:2019xhz}.

\begin{table}[b!]
    \centering
    \begin{small}
        \begin{tabular}{@{\hspace{2pt}}l|ll|ll@{\hspace{2pt}}}
            \toprule
            & \multicolumn{2}{c|}{toy model in \cref{sec:toy}}
            & \multicolumn{2}{c}{jet tagging in \cref{sec:jets}} \\
            LMC hyperparameter
            & latent chain & feature chain
            & latent chain & feature chain \\
            \midrule
            \# steps $\tau$         & $50$              & $100$             & $30$              & $30$ \\
            step size $\lambda$     & $\num{5e-3}$    & $\num{5e-3}$    & $\num{7.2e-5}$    & $\num{3.6e-5}$ \\
            noise scale $\sigma$    & $\num{1e-1}$    & $\num{1e-1}$    & $\num{1e-2}$    & $\num{1e-4}$ \\
            gradient clipping       & \xmark            & \xmark            & $\num{1e-2}$    & $\num{1e-2}$ \\
            sample clipping         & \xmark            & $[-4.5, 4.5]$     & \xmark            & $[0.0, 1.0]$ \\
            boundary rejection      & \xmark            & \cmark            & \xmark            & \xmark \\
            Metropolis-adjusted     & \cmark            & \cmark            & \cmark            & \cmark \\
            noise annealing         & \xmark            & \xmark            & \xmark            & \cmark \\
            \bottomrule
        \end{tabular}
    \end{small}
    \caption{LMC sampling parameters in latent and feature space.}
    \label{tab:hyperparameters_lmc}
\end{table}

\subsubsection*{Sampling hyperparameters}

The (B)NAE training is a significant bottleneck because the energy-based objective requires samples from the model distribution $\ptheta(x)$ to evaluate the gradient in Eq.\eqref{eq:training_ebm}~\cite{yoonAutoencodingNormalizationConstraints2023,Dillon:2022mkq}. As discussed in \cref{sec:bnae}, we generate these samples with Langevin Monte Carlo (LMC), using the Metropolis-adjusted Langevin algorithm (MALA). To improve sampling efficiency, we use the on-manifold initialization (OMI) strategy of Ref.~\cite{yoonAutoencodingNormalizationConstraints2023}. OMI first evolves a Markov chain in latent space and maps its endpoint through the decoder to initialize the feature-space chain. 

We initialize the latent chain from a replay buffer with probability $0.95$ and from a noise distribution with probability $0.05$. The replay buffer contains $\num{10000}$ endpoints from previous latent chains and reduces the number of LMC steps required during (B)NAE training~\cite{yoonAutoencodingNormalizationConstraints2023}. For the toy model, where $z \in \real^{d_{z}}$, we initialize noise from a standard normal distribution. For jet tagging we initialize a standard Gaussian vector in $\real^{d_{z}}$ and normalize it, $z \leftarrow z/\vert z\vert$, such that the resulting latent vectors are uniformly distributed over the unit hypersphere $\mathbb{S}^{d_{z}-1}$. During jet-tagging training, we normalize the latent vector after each LMC step to project it back onto the hypersphere. For jet tagging, in the feature-space chain, we anneal the stochastic noise according to $\sigma_{t} = \sigma_{0}/(1+t)$, with $t=0,\ldots,\tau-1$, so that the chain explores more broadly at early steps and performs smaller updates at later steps. We additionally clip the energy gradients in both latent and feature space at $0.01$ at each LMC step to improve sampling stability. We use the same temperature for the latent- and feature-space chains, $\tilde{T} = T$. For the toy model, we choose the LMC step size $\lambda$ and noise scale $\sigma$ according to $2\lambda = \sigma^{2}$. For jet tagging, we instead choose $2\lambda>\sigma^2$, following Ref.~\cite{Dillon:2022mkq}, to increase the relative contribution of the gradient drift. From Eq.\eqref{eq:eff_temperature}, this choice corresponds to an effective temperature $T_{\text{eff}} < T$ and accelerates convergence of the LMC chain. 

Following Ref.~\cite{yoonAutoencodingNormalizationConstraints2023}, for both the toy model and jet tagging, we generate samples from $\ptheta(x)$ using latent chains that are eight times longer than the chains we use during training, with $\tilde{\tau} = 400$ for the toy model and $\tilde{\tau} = 240$ for jets, and without using the replay buffer.

We summarize the latent- and feature-space LMC hyperparameters for the toy model and for jet tagging in \cref{tab:hyperparameters_lmc}. 

\subsubsection*{Architecture and training}
\label{app:network_training}

For the BNAEs, we use the Bayesian layers of Ref.~\cite{shridhar2019comprehensive} with a mean-field variational posterior, where each network parameter follows an independent Gaussian distribution~\cite{Graves:2011varinf,Blundell:2015bay}. With a Gaussian prior, we evaluate the KL divergence in Eq.\eqref{eq:bnae_loss} analytically. We set the prior standard deviation to $1.0$ for the Bayesian linear layer (toy model) as well as for the Bayesian convolutional layer (jet tagging). During training, we estimate the posterior expectation with a single Monte Carlo sample and use the local reparameterization trick to reduce the variance of the stochastic gradients~\cite{Kingma:2015vardrop}. For the BNAE, we replace the NAE's sigmoid output activation with a linear activation, since the sigmoid led to unstable BAE training.

For the toy model, we use the default (B)NAE configuration summarized in \cref{tab:hyperparameters_toy}, closely following Ref.~\cite{yoonAutoencodingNormalizationConstraints2023}. Both the encoder and decoder are multilayer perceptrons (MLPs) with three hidden layers of $128$ units each. Since the toy model uses a Euclidean latent space, $z \in \real^{d_{z}}$, we include a latent-space regularization term proportional to $\vert z\vert^{2}$, which favors latent representations near the origin.

For jet tagging, we use the same architecture and hyperparameters for the top- and QCD-tagging tasks to maintain a symmetric anomaly-detection setup. We summarize the corresponding configuration in \cref{tab:hyperparameters_jets}. Our setup largely follows Ref.~\cite{Dillon:2022mkq}. Since we train with $T \neq 1$, we rescale the negative-energy regularization to keep the different contributions to the loss at comparable scales. We use the convolutional autoencoder architecture of Ref.~\cite{Dillon:2022mkq}, summarized in \cref{tab:architecture}. The network contains five convolutional layers, four fully connected layers, and four transposed-convolutional layers. We use PReLU activations throughout the hidden layers, a linear activation at the encoder output, and a sigmoid activation at the decoder output. Following Ref.~\cite{Dillon:2022mkq}, we apply spectral normalization~\cite{Miyato:2018tgo} to every NAE layer. We further stabilize NAE training with two regularization terms. We apply an L2 penalty to the encoder and decoder weights with coefficient $10^{-8}$. We also introduce a negative-energy regularization term, defined as the mean squared energy of the negative samples, with coefficient $\num{2.5e4}$ to prevent the negative-sample energies from diverging. We do not apply latent-space regularization for jet tagging. For the BNAE, we keep the negative-energy regularization but omit both the L2 weight penalty and spectral normalization. 

\begin{table}[H]
    \centering
    \begin{small}
        \begin{tabular}{@{\hspace{2pt}}l|l@{\hspace{2pt}}}
            \toprule
            hyperparameter & NAE and BNAE network \\
            \midrule
            network optimizer & \adam algorithm~\cite{Kingma:2014vow} with default parameters \\
                      & ($\beta_1=0.9$, $\beta_2=0.999$, $\epsilon=10^{-8}$, weight decay $\lambda=0$) \\
            learning rate & AE $10^{-3}$ and NAE $10^{-5}$ (constant, no LR scheduler) \\
            batch size & $2048$ \\
            \# epochs & AE $100$ and NAE $100$ \\
            \midrule
            temperature $T=\tilde{T}$ & $0.2$ (initial), trainable \\
            temperature optimizer & \adam~\cite{Kingma:2014vow}, default parameters, constant LR $10^{-3}$ \\
            replay buffer size & $\num{10000}$ \\
            replay ratio & $0.95$ \\
            latent space & Euclidean $\real^{d_{z}}$ with $d_{z} = 3$ \\
            inner activations & Rectified Linear Unit (ReLU) \\
            output activation & linear \\
            \midrule
            layer regularization & none \\
            latent regularization & $\num{7.069e-6}$ \\
            L2 weight regularization & none \\
            negative-energy regularization & $1.0$ \\
            \bottomrule
        \end{tabular}
    \end{small}
    \caption{Default (B)NAE setup used for the toy model in \cref{sec:toy}.}
    \label{tab:hyperparameters_toy}
\end{table}

\begin{table}[H]
	\centering
	\begin{small}
		\begin{tabular}{@{\hspace{2pt}}l|l@{\hspace{2pt}}}
			\toprule
            hyperparameter & NAE and BNAE network \\
            \midrule
			optimizer & \adam algorithm~\cite{Kingma:2014vow} with default parameters \\
            & ($\beta_{1}=0.9$, $\beta_{2}=0.999$, $\epsilon=10^{-8}$, weight decay $\lambda=0$) \\
            learning rate & AE $10^{-3}$ and NAE $10^{-5}$ (constant, no LR scheduler) \\
            batch size & AE $2048$ and NAE $128$ \\
            \# epochs & AE $500$ and NAE $100$ \\
            \midrule
            temperature $T=\tilde{T}$ & $\num{6e-7}$ (not trainable) \\
            replay buffer size & $\num{10000}$ \\
            replay ratio & $0.95$ \\
            latent space & spherical $\mathbb{S}^{d_{z}-1}$ with $d_{z} = 6$\\
            inner activations & parametric rectified linear unit (PReLU) \\
            output activation & sigmoid (NAE) and linear (BNAE) \\
            \midrule
            layer regularization & spectral normalization~\cite{Miyato:2018tgo} (NAE) and none (BNAE) \\
            latent regularization & none \\
            L2 weight regularization & $10^{-8}$ (NAE) and none (BNAE) \\
            negative-energy regularization & $\num{2.5e4}$ \\
			\bottomrule
		\end{tabular}
	\end{small}
    \caption{Default (B)NAE setup used for jet tagging in \cref{sec:jets}.}
	\label{tab:hyperparameters_jets}
\end{table}

\begin{table}[H]
	\centering
	\begin{small}
		\begin{tabular}{@{\hspace{2pt}}l|l@{\hspace{2pt}}}
			\toprule
			\parbox[t]{10pt}{\multirow{7}{*}{\rotatebox[origin=c]{90}{encoder}}}
			& Conv2d(1, 8, 3, 1, 1, True) --- PReLU --- \\
			& Conv2d(8, 8, 3, 1, 1, True) --- PReLU --- MaxPool2d(2, 2) --- \\
			& Conv2d(8, 8, 3, 1, 1, True) --- PReLU --- \\
			& Conv2d(8, 8, 3, 1, 1, True) --- PReLU --- \\
			& Conv2d(8, 1, 3, 1, 1, True) --- PReLU --- Flatten --- \\
			& Dense(400, 100, True) --- PReLU --- \\
			& Dense(100, $d_{z}$, True) --- normalization to hypersphere $\mathbb{S}^{d_{z}-1}$ \\
			\midrule
			\parbox[t]{10pt}{\multirow{6}{*}{\rotatebox[origin=c]{90}{decoder}}}
			& Dense($d_{z}$, 100, True) --- PReLU --- \\
			& Dense(100, 400, True) --- PReLU --- Reshape(1, 20, 20) --- \\
			& Deconv2d(1, 8, 3, 1, 1, True) --- PReLU --- \\
			& Deconv2d(8, 8, 3, 1, 1, True) --- PReLU --- Upsampling(2, ``bilinear'') --- \\
			& Deconv2d(8, 8, 3, 1, 1, True) --- PReLU --- \\
			& Deconv2d(8, 1, 3, 1, 1, True) --- sigmoid (NAE) or linear (BNAE) \\
			\bottomrule
		\end{tabular}
	\end{small}
    \caption{Convolutional autoencoder architecture taken from Ref.~\cite{Dillon:2022mkq}.}
    \label{tab:architecture}
\end{table}

\clearpage
\section{Supplementary results}
\label{app:results}

\subsection{Toy model uncertainty calibration}

\begin{figure}[H]
    \includegraphics[page=14, width=0.45\linewidth]{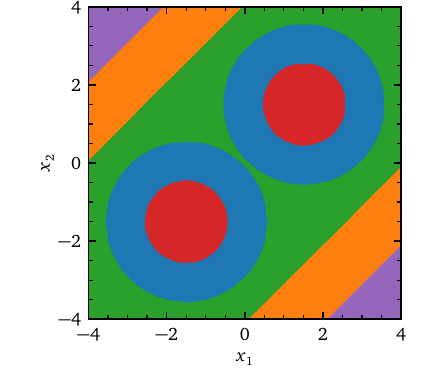}\hfill
    \includegraphics[page=17, width=0.45\linewidth]{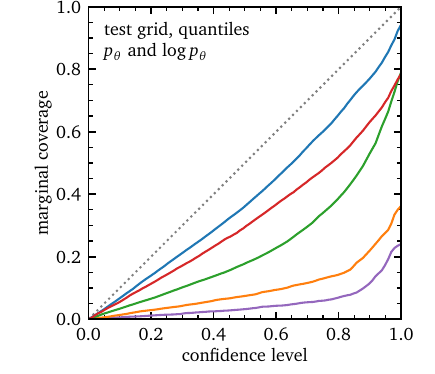}\\
    \includegraphics[page=15, width=0.45\linewidth]{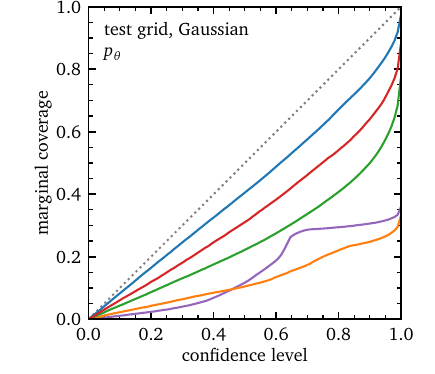}\hfill
    \includegraphics[page=16, width=0.45\linewidth]{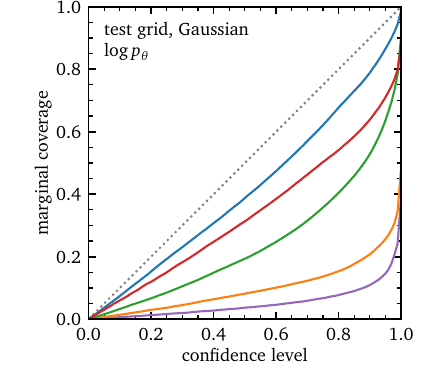}
    \caption{Marginal coverages of the two-dimensional toy model, evaluated for different phase-space regions of the test grid. For the coverages, we show the quantile-based construction of Eq.\eqref{eq:quantiles} and the Gaussian intervals from Eq.\eqref{eq:gaussian}.}
    \label{fig:coverage_phase}
\end{figure}

We assess the calibration of the BNAE uncertainty estimates using coverage tests over the two-dimensional $\dgrid$. In the upper-left panel of \cref{fig:coverage_phase}, we define the corresponding phase-space areas, and in the remaining panels we show the marginal coverage as a function of the nominal confidence level. For both $\ptheta(x)$ and $\log\ptheta(x)$, all coverage curves fall systematically below the diagonal, demonstrating under-coverage across $\dgrid$. We generally observe stronger under-coverage away from the high-density modes, particularly in the green, orange, and purple regions, where the BNAE receives less training support. However, the high-density red region shows stronger under-coverage than the intermediate blue region. The quantile-based estimates for $\ptheta(x)$ and $\log\ptheta(x)$ show similar coverage behavior. For $\log\ptheta(x)$, the Gaussian-based estimates closely follow the quantile-based results across all phase-space regions. For $\ptheta(x)$, the Gaussian-based estimates preserve the same ordering and overall level of under-coverage, with only small differences in the regions with the lowest coverage.

\clearpage
\subsection{(B)NAE-generated high-level features}
\label{app:nae_generated_samples}

Because low-level jet images are difficult to compare directly, we compare (B)NAE-generated samples and truth samples using standard high-level jet-substructure observables~\cite{Marzani:2019hun,Bonilla:2022wzp,Kasieczka:2018lwf,Butter:2022xyj,Vent:2025ddm}. We consider the jet invariant mass, $m_{\text{jet}}$, which we compute by reconstructing each constituent four-momentum from the normalized pixel intensity as $\pt$ and the corresponding pixel coordinates as $\eta$ and $\phi$. To probe the jet's prong structure, we use $N$-subjettiness~\cite{Thaler:2010tr,Thaler:2011gf,Larkoski:2013eya},
\begin{align}
    \tau_{N}^{(\beta)}
    = \min_{\{\hat{n}_{k}\}}
    \frac{1}{d_{0}} \sum_{i\in\text{jet}}
    \pti \min_{k=1,\ldots,N} (\Delta R_{ik})^{\beta}
    \qquad\mwith\quad
    d_{0} = \sum_{i\in\text{jet}} \pti R_{0}^{\beta} \eqc
    \label{eq:nsubjettiness}
\end{align}
where $\hat{n}_{k}$ denotes the $k$-th subjet axis and $\Delta R_{ik}$ the angular distance between constituent $i$ and that axis. Smaller $\tau_{N}$ indicates greater compatibility with an $N$-prong radiation pattern. We set $\beta = 1$ throughout and determine the subjet axes using exclusive-$\kt$ clustering~\cite{Thaler:2010tr}.

Ratios of consecutive $N$-subjettiness observables, $\tau_{N}/\tau_{N-1}$, provide powerful discriminants of the prong structure of a jet. In particular, $\tau_{2}/\tau_{1}$ probes two-prong decays, while $\tau_{3}/\tau_{2}$ probes the characteristic three-prong structure of boosted top jets~\cite{Plehn:2009rk}. For visualization, however, we show the individual $\tau_{N}$ distributions rather than their ratios, since the ratio can obscure differences between the numerator and denominator distributions. 

\begin{figure}[H]
    \includegraphics[page=01, width=0.45\linewidth]{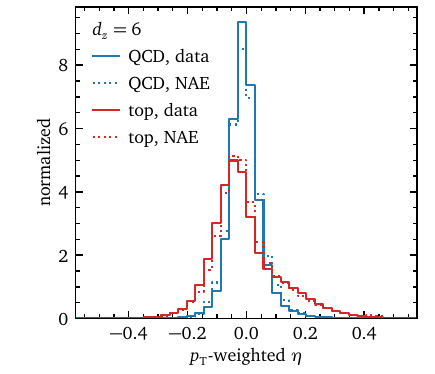}\hfill
    \includegraphics[page=02, width=0.45\linewidth]{figures/samples_nae_6d.pdf}\\
    \includegraphics[page=03, width=0.45\linewidth]{figures/samples_nae_6d.pdf}\hfill
    \includegraphics[page=04, width=0.45\linewidth]{figures/samples_nae_6d.pdf}\\
    \includegraphics[page=05, width=0.45\linewidth]{figures/samples_nae_6d.pdf}\hfill
    \includegraphics[page=06, width=0.45\linewidth]{figures/samples_nae_6d.pdf}
    \caption{NAE with a six-dimensional latent space. The panels compare true and NAE-generated jets.}
    \label{fig:6d_hlf}
\end{figure}

In \cref{fig:6d_hlf}, we show the $\pt$-weighted $\eta$ and $\phi$ (top panels), $m_{\text{jet}}$ (center left panel), $\tau_{1}$ (center right panel), $\tau_{2}$ (bottom left panel), and $\tau_{3}$ (bottom right panel) for QCD jets (blue) and top jets (red), comparing NAE-generated samples (dashed) with the corresponding truth samples (solid). The agreement is strongest for the angular distributions and remains good for $m_{\text{jet}}$, $\tau_{1}$, and $\tau_{2}$. For top jets, the NAE underestimates the peak and left tail of both $m_{\text{jet}}$ and $\tau_{1}$. For QCD jets, it slightly reshapes the peaks of $\tau_{2}$ and $\tau_{3}$ and overestimates their high-value tails. The largest discrepancy appears in the top-jet $\tau_{3}$, where the NAE shifts the distribution toward larger values and produces a broader tail. Overall, the NAE reproduces the high-level observables reasonably well but has more difficulty capturing the three-prong structure that $\tau_{3}$ encodes.

\begin{figure}[H]
    \includegraphics[page=01, width=0.45\linewidth]{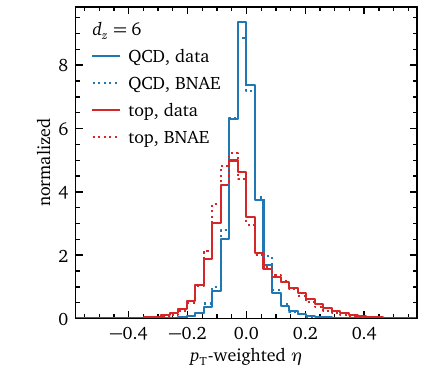}\hfill
    \includegraphics[page=02, width=0.45\linewidth]{figures/samples_bnae_6d.pdf}\\
    \includegraphics[page=03, width=0.45\linewidth]{figures/samples_bnae_6d.pdf}\hfill
    \includegraphics[page=04, width=0.45\linewidth]{figures/samples_bnae_6d.pdf}\\
    \includegraphics[page=05, width=0.45\linewidth]{figures/samples_bnae_6d.pdf}\hfill
    \includegraphics[page=06, width=0.45\linewidth]{figures/samples_bnae_6d.pdf}
    \caption{BNAE with a six-dimensional latent space. The panels compare true and BNAE-generated jets.}
    \label{fig:6d_hlf_bnae}
\end{figure}

For the BNAE with $d_{z}=6$, the generated samples show a similar level of agreement as seen in \cref{fig:6d_hlf_bnae}. The BNAE reproduces the $\pt$-weighted $\eta$ and $\phi$ distributions well for both QCD and top jets. It also captures the overall shapes of $m_{\text{jet}}$, $\tau_{1}$, and $\tau_{2}$, although it performs somewhat worse than the NAE. In particular, the BNAE shifts the top-jet $m_{\text{jet}}$ and $\tau_{1}$ distributions toward smaller values and underestimates their peaks. For QCD jets, it slightly distorts the low-value regions of $m_{\text{jet}}$, $\tau_{1}$, and $\tau_{2}$. The largest discrepancy again appears in $\tau_{3}$. The BNAE shifts the top-jet distribution toward smaller values, sharpens the peak, and underestimates the high-$\tau_3$ tail. It also modifies the QCD $\tau_{3}$ peak and tail.

\begin{figure}[H]
    \includegraphics[page=01, width=0.45\linewidth]{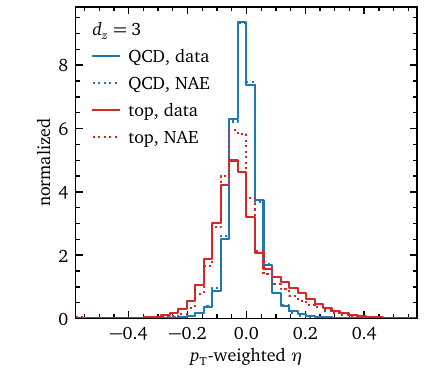}\hfill
    \includegraphics[page=02, width=0.45\linewidth]{figures/samples_nae_3d.pdf}\\
    \includegraphics[page=03, width=0.45\linewidth]{figures/samples_nae_3d.pdf}\hfill
    \includegraphics[page=04, width=0.45\linewidth]{figures/samples_nae_3d.pdf}\\
    \includegraphics[page=05, width=0.45\linewidth]{figures/samples_nae_3d.pdf}\hfill
    \includegraphics[page=06, width=0.45\linewidth]{figures/samples_nae_3d.pdf}
    \caption{NAE with a three-dimensional latent space. The panels compare true and NAE-generated jets.}
    \label{fig:3d_hlf}
\end{figure}

Ref.~\cite{Dillon:2022mkq} used an NAE with latent dimension $d_{z}=3$, which degrades the distributions significantly as we show in \cref{fig:3d_hlf}. The NAE-generated samples reproduce the angular distributions somewhat well, but the NAE shows larger deviations in the jet-substructure observables. It broadens the QCD $m_{\text{jet}}$ distribution and underestimates the top-jet mass peak. It also distorts the $\tau_{1}$ and $\tau_{2}$ distributions, especially for top jets, where it shifts probability toward smaller values and suppresses the peaks. The largest deviations again occur in $\tau_{3}$, where the NAE produces pronounced shape distortions and additional peak structure for both QCD and top jets. 

\clearpage
\subsection{BNAE-learned anomaly score with uncertainties}
\label{app:bnae_energies}

\begin{figure}[H]
    \includegraphics[page=04, width=0.45\linewidth]{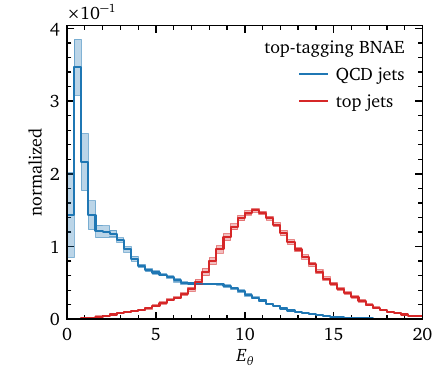}\hfill
    \includegraphics[page=05, width=0.45\linewidth]{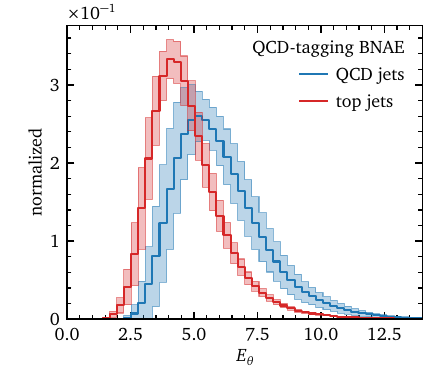}
	\caption{Energy distributions after training the BNAE on QCD jets (left) and on top jets (right). We show the energy for QCD jets (blue) and top jets (red) in both cases, together with their $\pm 1\sigma$ uncertainty bands from $50$ Monte Carlo samples.}
	\label{fig:jets_bnae_energies}
\end{figure}

In \cref{fig:jets_bnae_energies}, we show the BNAE energy distributions, which we use as anomaly scores, for top tagging (left) and QCD tagging (right). We show QCD jets in blue and top jets in red, together with the BNAE uncertainties. The BNAE assigns higher energies to top jets for top tagging and to QCD jets for QCD tagging, as expected when the energy tracks the learned density. As discussed in \cref{sec:jets_performance}, the BNAE shows larger uncertainties for QCD tagging than for top tagging. Compared with the NAE results in Ref.~\cite{Dillon:2022mkq}, the QCD-tagging distributions show less separation, although their overall ordering is consistent. By shifting QCD jets toward higher energies in the QCD-tagging direction, the BNAE goes beyond the simple compressibility ordering of a standard BAE.

\clearpage
\bibliographystyle{tepml}
\bibliography{tilman,references}

\end{document}